# Automated Dataset Generation System for Collaborative Research of Cyber Threat Analysis

Daegeon Kim  and Huy Kang Kim

Graduate School of Information Security, Korea University, Seoul, Republic of Korea

Correspondence should be addressed to Huy Kang Kim; cenda@korea.ac.kr

## Abstract

The objectives of cyberattacks are becoming sophisticated, and attackers are concealing their identity by masquerading as other attackers. Cyber threat intelligence (CTI) is gaining attention as a way to collect meaningful knowledge to better understand the intention of an attacker and eventually predict future attacks. A systemic threat analysis based on data acquired from actual cyber incidents is a useful approach to generating intelligence for such an objective. Developing an analysis technique requires a high volume and fine quality data. However, researchers can become discouraged by an inaccessibility to data because organizations rarely release their data to the research community. Owing to a data inaccessibility issue, academic research tends to be biased toward techniques that develope steps of the CTI process other than analysis and production.

In this paper, we propose an automated dataset generation system called CTIMiner. The system collects threat data from publicly available security reports and malware repositories. The data are stored in a structured format. We released the source codes and dataset to the public, including approximately 640,000 records from 612 security reports published from January 2008 to June 2019. In addition, we present a statistical feature of the dataset and techniques that can be developed using it. Moreover, we demonstrate an application example of the dataset that analyzes the correlation and characteristics of an incident. We believe our dataset will promote collaborative research on threat analysis for the generation of CTI.

## 1. Introduction

Cyber threat intelligence (CTI) is evidence-based knowledge including context, mechanisms, indicators, implications, and actionable advice regarding existing or emerging threats to assets [1]. CTI can be utilized to achieve a broad situational awareness, collaborate in defeating cyber threats faced by others, and prevent cyber threats by applying CTI into defense systems.

With an increase in global cyber threats, CTI is gaining increased attention as a response to such threats. Many nations and organizations have also attempted to promote the use of CTI by enacting laws that legalize and encourage the collection of CTI [2], sharing CTI through multilateral cooperation [3-5], and establishing various standards [6, 7]. Furthermore, during the recent decade, the number of articles related to CTI have dramatically increased, as shown in Figure 1[1].

---

[1] Google Scholar search result with exact keyword matching of 'cyber threat intelligence' including patents and citations on March 30, 2019.

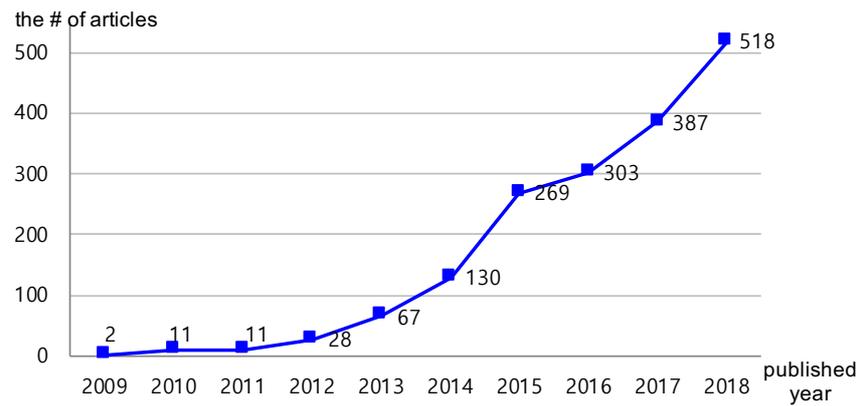

Figure 1: Number of articles related to CTI within the most recent decade

During the Olympic Winter Games in PyeongChang 2018, a cyberattack targeting the server operated by the organizing committee occurred. What makes this case noteworthy is that the security researchers attributed different countries as the perpetrators of the attack. The authors in [4] and [8] insisted that Chinese and Russian actors were responsible for the attack, respectively. In [9] and [10], it was pointed out that it is impossible to attribute the attack to a specific country based on the small amount of code discovered, which overlaps malware used by the Lazarus Group, a hacking group from North Korea. However, the authors of [4] insisted that there was evidence indicating that a Russian attacker tried to masquerade as a North Korean hacking group. In this example, we can see that a precise evidence-based analysis that considers all possibly related cases is vitally important for CTI generation.

However, among the traditional intelligence processes [11], i.e., planning and direction, collection, processing and exploitation, analysis and production, and dissemination and integration, most technical studies on CTI have tended to focus on steps other than analysis and production, which require a real CTI dataset. Despite the many advantages of a CTI analysis as mentioned in [12], such as 1) an interoperability of the data (machine, vendor, and organization independent), 2) a compact expression of the heterogeneous source of the threat information, and 3) the possibility of conducting a long-term and nation-wide threat analysis, we believe that the most challenging aspect of such a study is the limited accessibility of data to researchers. Although some web services provide functionality when searching for threat data, they do not offer a sufficient and useful set of data for research purposes. In addition, most of the datasets consist of only specific data types, e.g., the IP, URL, or hash value, and some datasets are strictly restricted to access in certain regions or to people of a particular nationality.

In this paper, we propose a cyber threat dataset generation system called CTIMiner, which automatically collects data from public security reports and malware repository websites and stores the data in a structured format. The generated dataset contains several types of data including malware analysis information, which consists of the file path, mutex, code sign information, and the other data types listed above. The main contributions of our work are as follows:

- Promoting collaborative CTI analysis research by proposing a cyber threat data generation system and a public database
- Demonstrating the use of the dataset for a correlation analysis
- Suggesting the development of techniques to generate CTI from a dataset

At this point, it would be warranted to introduce the techniques used to generate CTI from a dataset. However, this is beyond the scope of this paper and remains as our future research concern. We believe that the suggestion of the required techniques for analyzing the dataset can inspire the researchers and promote research into CTI analysis.

The remainder of this paper is organized as follows. The intelligence process and its associations with CTI activities are presented in section 2 with several studies related to each step. The overall system architecture of CTIMiner and the phases composing the run-time process are described in section 3. The dataset structure, the data categories, and the statistical features are detailed in section 4. In section 5, the dataset usage is demonstrated, and analysis techniques are suggested. Following an introduction of the source code and dataset access in section 6, we provide some concluding remarks in section 7.

## 2. Related Works

### 2.1. Intelligence Process and Automated CTI Activities

In the field of military operation, the well-defined intelligence process illustrated in Figure 2 was adapted to efficiently generate intelligence from low-level data collected in the field to support the decision-making process. This process is intended to be followed by a human intelligence officer but can also be projected into automated CTI activities.

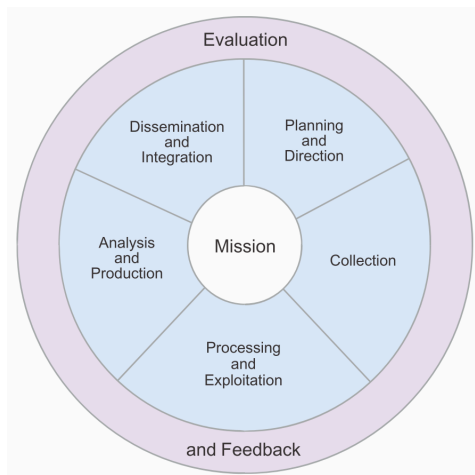

Figure 2: Intelligence cycle defined during military operation [11]

Once the operation direction is determined to fulfill the identified intelligence requirement, the raw data are collected and extracted from the sensors and data sources, which have the ability and functionality to obtain such data. The data gathered from these various sources are combined and converted into forms, in other words information, allowing the data to be efficiently analyzed. The information is passed into an analysis algorithm, such as a big data or machine learning based method, which enables the intelligence collected to be used by human analysts. Such intelligence is then spread to others who have access to it. The shared intelligence can also be integrated into the intelligence already available to users.

The association between the intelligence process and automated CTI activities is illustrated in Figure 3. In the following subsections, previous studies regarding CTI are introduced in order of the applied intelligence process, excluding the planning and direction steps because these are more strategic, rather than technical, concerns.

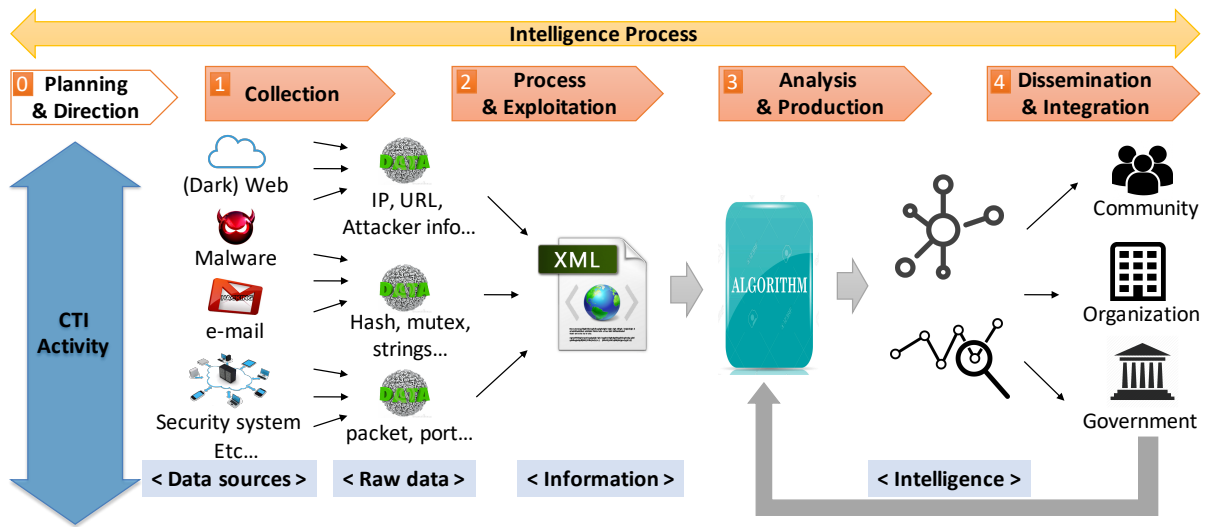

Figure 3: Association between the intelligence process and automated CTI activities

## 2.2. Collection

Because CTI is also a product of threat data processing through the intelligence process, low-level threat data can be collected during this step. Goel classified the types of data to be collected into unstructured and network data [13]. The former typically consists of hacker forum postings, blogs, and websites, whereas the latter is generated from information security systems such as firewalls, intrusion detection systems, and honeynets.

Benjamin et al. proposed a method for extracting information from hacker forums, IRC channels, and carding shops to identify threats [14]. In addition, Fachkha and Debbabi characterized the darknet and compared several methods for extracting threat information there [15].

As a data repository for research regarding cyber security analysis, the Information Marketplace for Policy and Analysis of Cyber-risk & Trust (IMPACT) [16], which is based on Protected Repository for the Defense of Infrastructure Against Cyber Threats (PREDICT) [17], provides several types of data, such as network flow, IDS and firewall, and unsolicited email data. It also provides useful tools for data analysis. However, the service is only available to DHS-approved countries, namely, the United States, Australia, Canada, Israel, Japan, Netherlands, Singapore, and the United Kingdom.

## 2.3. Processing and Exploitation

During the processing and exploitation step, raw data collected are converted into forms that can be readily applied by intelligence analysts and other users. Unstructured data and heterogeneous sources of data having different structures can be stored in a unified data format during this step for further analysis.

STIX [18] and OpenIOC [19] proposed by MITRE and MANDIANT are representative standards for expressing threat data. Specifically, STIX is widely used owing to the scalability of its schema, which uses components such as CybOX, and CAPEC. Liao et al. proposed an element extraction method for constructing structured data from unstructured data [20]. One notable aspect of this approach is that the meaning of the elements in the context can also be retrieved using a natural language processing technique.

## 2.4. Analysis and Production

During the analysis and production step, all processed information is integrated, evaluated, analyzed, and interpreted to produce intelligence. Kornmaier and Jaouën insisted that, to generate operational or strategic intelligence beyond tactical information, which is technical in nature, the threat data should be fused with data collected from different disciplines such as Human Based Intelligence (HUMINT), Imagery Intelligence (IMINT), Signal Intelligence (SIGINT), and Geographic Intelligence (GeoINT) [21].

Modi et al. proposed an automated threat data fusing system that correlates data crawled from the web by applying a string-matching based approach [22]. Similar commercial CTI services have also been developed, such as iDefense® IntelGraph by Verisign and a web intelligence engine by Recorded Future that allows users to navigate through extensive threat data following a string-matching correlation. One key feature of Recorded Future is that it can conduct a predictive analysis of specific future events through the use of information compiled in advance [23]. However, commercial services provide an indicator-centric analysis approach making it difficult to trace the correlation between incidents.
Kim et al. proposed a general framework for an efficient CTI correlation analysis by adopting a novel concept that expresses similarity between threat events in a graphical structure [12]. A graphical structure allows the analysts to trace the specifications and transition of related cyber incidents to infer an attacker's intention.

Using a threat report as the source of information, Qamar et al. proposed an automated mechanism to analyze the risk of a reported threat toward a networked system [24]. For this purpose, they defined the ontology of the IoCs, network, associated risk, and their relations. For the risk analysis of a networked system, four parameters, namely, the threat relevance, threat likelihood, total loss of affected assets, and threat reachability, are defined.

## 2.5. Dissemination and Integration

During the dissemination and integration step, intelligence is delivered to and used by the consumer. A guideline [25] and technical standard protocol [26] have been developed for sharing CTI. In addition, MISP [27], MANTIS [28], and CIF [29] are useful open-source platforms to store and share CTI.

As more participants in a community share CTI, access control issues with the shared data often arise. Zhao and White proposed an access control model that extends the group-centric Secure Information Sharing model to support collaborative information sharing in a community [30]. Although such assistive technologies promote CTI sharing, for example, social and political issues, the authority to operate CTI sharing policies and the trust management within a community are often controversial when establishing collaborative CTI sharing.

## 2.6. Data, Information, and Intelligence

In many CTI-related studies, the terms, *data*, *information*, and *intelligence*, are often intermixed without clarification. We need to use them clearly, as illustrated in Figure 4, based on the definition in [11].

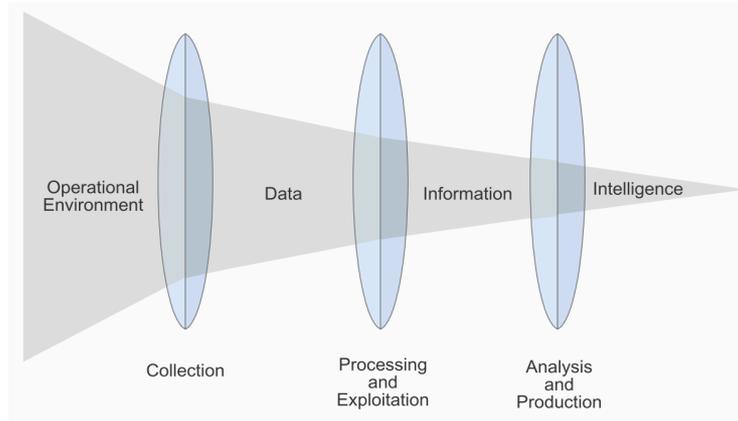

Figure 4: Data, information, and intelligence

*Data* are the individual facts collected from sensors in an operational environment. *Information* is data gathered and processed into an intelligible form, and *intelligence* is the new understanding of current and past information that allows a prediction of the future and informed decisions.

These definitions are applied not only to the general intelligence process but also to CTI activities. Throughout the data fusion and mining process, Bass defined data as measurements and observations; information as data placed in context, indexed, and organized; and knowledge, which is equal to intelligence, as information that has been explained or understood [30].

## 3. CTIMiner System Architecture

We propose a cyber threat data collecting system, CTIMiner, using the system architecture presented in Figure 5. The CTI collecting procedure is composed of three phases. During the first phase, the system gathers threat data from publicly accessible cyber intelligence reports published by organizations and companies. It also collects additional related data from a malware repository during the second phase. Finally, all collected data are stored in the database after passing through the last phase generating combined information in a structured format.

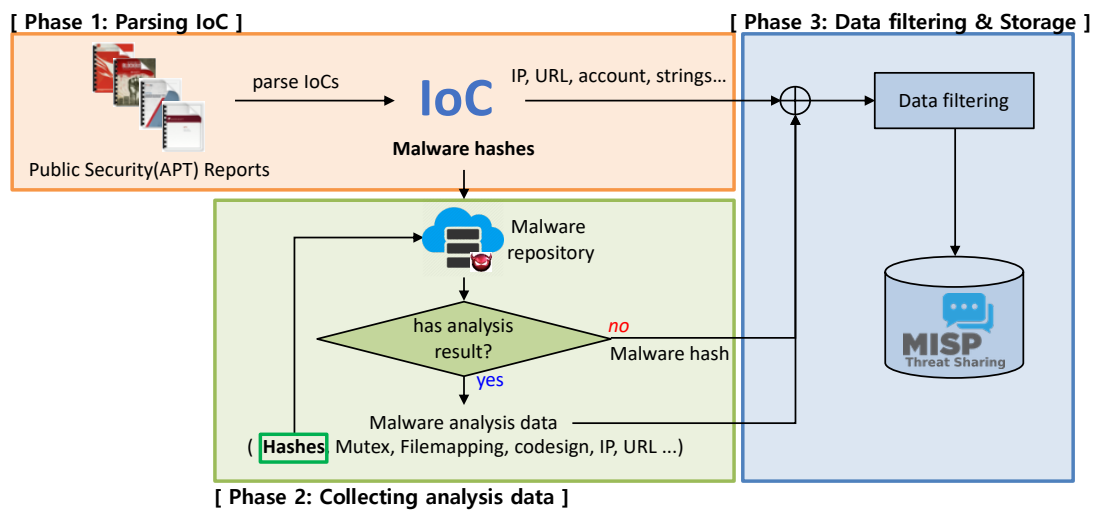

Figure 5: CTIMiner system architecture

## 3.1. Phase 1: Parsing Indicators of Compromise

This phase starts with collecting cyber intelligence reports that analyze cyber incidents and malware interrelated APT campaigns and groups. For this, we obtain a list of papers from public repositories[2] that provide publicly available articles and blog content related to malicious attacks, activities, and software associated with vendor-defined APT groups and/or tool sets. To maintain the usability of the dataset, we exclude the periodically published threat analysis reports from a list integrating the analysis results from different APT groups that have no interrelation with each other. Therefore, it can be assumed that the data extracted during phases 1 and 2 are related to the same (or related) threat actors. We can use this property to set the ground truth of the data for analysis. This property and the dataset usability are explained in detail in sections 4 and 5, respectively.

Next, Indicators of Compromise (IoCs) are extracted from the reports using a parser. We utilize a modified ioc_parser[3] that extracts IoCs matched by predefined regular expressions such as the URL, host, IP address, e-mail account, hashes (MD5, SHA1, and SHA256), common vulnerabilities and exposures (CVE), registry, file names ending with specific extensions, and the program database (PDB) path. Among the data obtained, the malware hash values are passed to the second phase for further data collection, and other values are passed to the last phase.

The IoC extraction performance is critically influenced by the performance of the parser. Therefore, other parsers can be chosen to increase the performance of this phase.

## 3.2. Phase 2: Collecting Analysis Data

Owing to the functional limitation of a parser, there may be remaining IoCs not extracted from the reports that can be found in malware analysis data. Moreover, we can obtain additional data from the analysis results that are not in the content of the reports. Notably, the valuable data, which cannot be expressed as a regular expression such as mutex, file mapping, code sign, and other strings, are only collectible from the malware analysis results.

To collect the malware analysis data, we use the malware repository service, malwares.com, operated by Saint Security Inc., the first cloud-based malware analysis platform in South Korea. It possesses over 800 million malware samples and maintains a partnership with VirusTotal. If the malware analysis results are retrieved by querying the hash value, the data in the results, namely, hashes, URLs, IP addresses, PDB paths, code signs, file names, and other strings, are passed to the last phase; otherwise, the hash value itself is passed. We do not store malware samples in the database because of the possible occurrence of copyright concerns when it is publicly released. For the new hash values found from the results, the analysis data are gathered through the same procedure.

## 3.3. Phase 3: Data Filtering and Storage

The data collected from several sources may be redundant or noisy and can be filtered out during this phase. For example, some files are automatically generated by the operating system regardless of the intent of the malware creator when the malware is executed. We merge the repetitive data and remove noisy data during this phase. However, the trade-off between false-positives and false-negatives needs to be considered for noise removal. The filtered data are stored in the MISP server, which provides an API to manage and export data in various structured formats.

---

[2] APTnotes (https://github.com/aptnotes/data)
  APT & CyberCriminal Campaign Collection (https://github.com/CyberMonitor/APT_CyberCriminal_Campaign_Collections)
[3] https://github.com/armbues/ioc_parser

Optionally, we categorize the data types composing the dataset and analyze their statistical characteristics during this phase, the results of which are presented in the next section.

### 3.4. System Processing Results of Phases 1 and 2

We ran this system on 612 collected APT reports published from January 2008 to June 2019, the numerical processing results of which are in Table 1. Among the 14,313 malware hashes extracted from the reports, we obtained analysis results for 68.1% of them from the malware repository. Among the analysis information, we found 450 new malware hashes that were not contained in the APT reports and added the analysis information to the dataset. The value of including the malware analysis data, in addition to the IoCs extracted from the reports, is described in the statistical analysis of the dataset provided in section 4.2.

Table 1: Processing results of phases 1 and 2

| Types | No. | % |
| --- | --- | --- |
| No. of reports | 612 | - |
| No. of data stored in the dataset | 642,810 | - |
| No. of extracted malware hashes from the reports | 14,313 | - |
| No. of analyzed malware | 9,753 | 68.1 |
| No. of additionally extracted malware | 450 | 4.6 |

## 4. Dataset Descriptions

### 4.1. Dataset Structure and Data Types

The dataset is composed of several sets of events, and Figure 6 shows the relationship of one set of events, which is composed of two types of events, namely, one report event and several malware events. A report event includes the data extracted from the first phase described in section 3, which parsed the texture IoCs from the APT reports. Malware events are created whenever malware hashes are detected, and it is possible to obtain their analyzed data during phase 2. These malware events and report events from which the malware hashes are originated can be grouped under the title of the report.

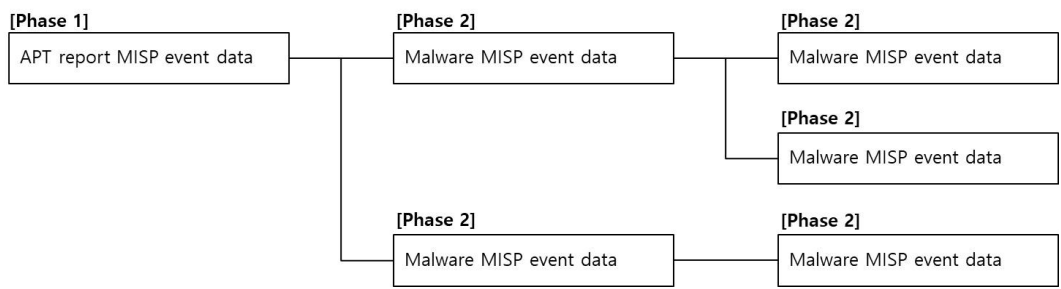

Figure 6: Relationship of a set of events

The data schema of an event is presented in Figure 7, and a short example of a set of events is shown in Figure 8. Because all malware events originating from one report include the same file name of the report, this can be used as the ground truth of the correlation analysis of the data. In addition, the malware compilation dates and the publication dates of the reports can be useful for a temporal analysis of the dataset. A sample application of the dataset for a correlation analysis using these dataset characteristics is demonstrated in section 5.

```
<Event>
    <id> the event ID automatically assigned by MISP </id>
    <date> the publication date of a report or the time stamp of malwre </date>
    <info> the title of the report or the hash value of malware </info>
    <Attribute>
        <item>
            <category> the category of the data </category>
            <comment> the comment about the data </comment>
            <value> the data value </value>
            <type> the data type </type>
            <id> the attribute ID automatically assigned by MISP </id>
        </item>
        <item>
            .
            .
            .
        <item>
</Event>
```

Figure 7: Data schema of an event

```
<Event>
    <id>2852</id>
    <date>2014-12-03</date>
    <info>Cylance_Operation_Cleaver_Report.pdf</info>
    <Attribute>
        <item>
            <category>External analysis</category>
            <comment />
            <value>zhcat.exe</value>
            <type>filename</type>
            <id>30996</id>
        </item>
        <item>
            <category>External analysis</category>
            <comment />
            <value>CVE-2010-0232</value>
            <type>vulnerability</type>
            <id>31056</id>
        </item>
        <item>
            <category>Network activity</category>
            <comment />
            <value>64.120.128.154</value>
            <type>ip-src</type>
            <id>30988</id>
        </item>
    </Attribute>
</Event>
```

```
<Event>
    <id>2741</id>
    <date>2014-12-03</date>
    <info>836ef6b06c5fd52ecc910a3e3408004a</info>
    <Attribute>
        <item>
            <category>External analysis</category>
            <comment>original_filename</comment>
            <value>zhCat.exe</value>
            <type>filename</type>
            <id>29863</id>
        </item>
        <item>
            <category>Network activity</category>
            <comment>TCP_undetected</comment>
            <value>1.224.181.13</value>
            <type>ip-src</type>
            <id>29864</id>
        </item>
        <item>
            <category>Payload installation</category>
            <comment />
            <value>723cdf97284e58a1672e031013620fe8d74e27f1</value>
            <type>sha1</type>
            <id>29861</id>
        </item>
        <item>
            <category>Artifacts dropped</category>
            <comment />
            <value>e:\Projects\Cleaver\trunk\MainModule\obj\Release\MainModule.pdb</value>
            <type>pdb</type>
            <id>29866</id>
        </item>
        <item>
            <category>Other</category>
            <comment />
            <value>Cylance_Operation_Cleaver_Report.pdf</value>
            <type>comment</type>
            <id>29867</id>
        </item>
    </Attribute>
</Event>
```

Figure 8: Example of a set of events

The types of attributes stored in a dataset are the IP, URL, e-mail address, date and time, CVE, file name, PDB path, digital code sign serial number, and other string data, including the author and title of the document. The amount of data, the report, and the malware events are shown in Table 2. Using the source codes that we publicly released, a dataset composed of the attribute types of interest can be created.

Table 2: Number of data for each type

| Year | Data Types | | | | | | | | | | | Report | Malware |
|------|------|--------|--------|--------|-----------|-----|-----------|-----|-----------|--------|---------|--------|---------|
| | Hash | IP | URL | e-mail | date, time | CVE | file name | PDB | code sign | others | total | | |
| 2008 | 0 | 3 | 171 | 0 | 0 | 0 | 17 | 0 | 0 | 0 | 191 | 2 | 0 |
| 2009 | 2 | 7 | 84 | 2 | 0 | 0 | 10 | 0 | 0 | 0 | 105 | 2 | 0 |
| 2010 | 223 | 79 | 280 | 14 | 32 | 2 | 213 | 0 | 0 | 0 | 800 | 7 | 32 |
| 2011 | 1,440 | 412 | 478 | 17 | 319 | 7 | 713 | 2 | 38 | 25 | 3,340 | 14 | 319 |
| 2012 | 2,240 | 433 | 637 | 46 | 465 | 30 | 828 | 2 | 43 | 7 | 4,524 | 22 | 465 |
| 2013 | 8,329 | 2,505 | 3,032 | 599 | 1,798 | 45 | 3,003 | 97 | 802 | 61 | 19,571 | 47 | 1,798 |
| 2014 | 5,614 | 5,484 | 3,282 | 476 | 1,116 | 83 | 2,804 | 22 | 438 | 28 | 18,842 | 100 | 1,116 |
| 2015 | 6,801 | 2,752 | 2,658 | 334 | 1,554 | 48 | 3,077 | 28 | 206 | 34 | 17,258 | 78 | 1,554 |
| 2016 | 8,001 | 525,020 | 3,449 | 235 | 1,833 | 81 | 4,873 | 43 | 154 | 14 | 543,703 | 79 | 1,974 |
| 2017 | 4,343 | 3,316 | 3,582 | 534 | 935 | 49 | 2,780 | 13 | 99 | 9 | 15,660 | 72 | 1,017 |
| 2018 | 3,900 | 3,296 | 2,582 | 229 | 0 | 74 | 2,660 | 34 | 404 | 31 | 13,210 | 125 | 1,300 |
| 2019 (–Jun.) | 2,046 | 719 | 1,439 | 194 | 0 | 51 | 1,110 | 9 | 36 | 2 | 5,606 | 64 | 628 |
| Total | 42,939 | 544,026 | 21,674 | 2,680 | 8,052 | 470 | 22,088 | 250 | 2,220 | 211 | 642,810 | 612 | 10,203 |

## 4.2. Data Categories and Statistics

We observed that the data collected from the reports and the malware analysis information are related to common cyber campaigns or threat actors, which can be categorized as shown in Figure 9.

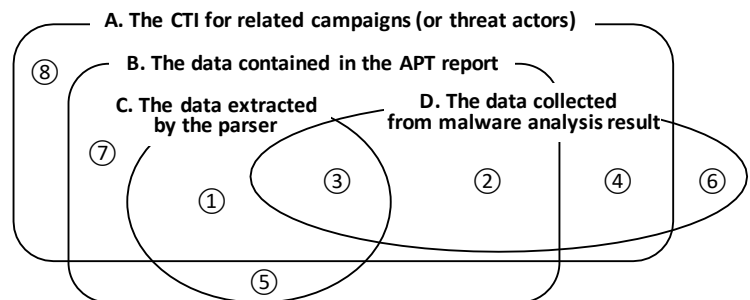

Figure 9: Data categories of the dataset

The characteristics of each category are as follows.

① Data that can only be extracted by the parser belong in this category. The quality and quantity of this type of data depend highly on the contents of the reports and the functionality of the parser.

② Malware analysis data contained in reports but unable to be extracted by the parser belong in this category. The volume of this type of data shows how much malware analysis data can compensate for the limitation of the parser. In addition, the indicator of this category can be used to compare the quality of the analysis results from several malware repositories.

③ This category includes the data extracted by the parser as well as by the malware analysis results.

④ Some data related to campaigns or threat actors can be excluded in the APT reports owing to the low priority compared to other information, or the analysis limitations of the authors. Such data found from malware analysis results belong to this category.

⑤ Noise data generated by the parser belong to this category. The functional limitation of the parser increases the portion of data in this category.

⑥ Data in this category are the noise generated from malware analysis information. It is difficult to distinguish between ④ and ⑥, but meaningless data generated by the runtime environment of malware belong to this category.

⑦ There are numerous data in the reports that are difficult for the parser or malware analysis information to obtain. Specifically, nontechnical information such as actors and groups of cyber campaigns mainly reside in this category. These data need to be extracted manually or by other supplemental methods.

⑧ This category is similar to ⑦ in the sense that neither data extraction methods can discover data in this category. Publishers can intentionally exclude the data, or may not even know about them. The volume of this category can be minimized by comparing several reports related to the same campaigns or threat actors, or by gathering multi-source information such as HUMINT and SIGINT.

The statistical features of these dataset categories generated through phase 3 of the system are listed in Table 3. It is worth noting that 43% of the data come from malware analysis results ( ② and ④) and 26% are newly discovered data that are not contained in the reports (④).

Comparing that the vast amount of data types in ② with the hash values, ④ consists of various types of data including code signs, IP addresses, and other string information valuable to identifying an incident.

Table 3: Percentage of data in each category

| Category | ① | ② | ③ | ④ |
|----------|-----|-----|-----|-----|
| % | 46 | 17 | 11 | 26 |

## 5 Dataset Application

As mentioned previously, the objective of generating our dataset is to promote academic research related to CTI analysis. We propose three research topics applying a dataset and demonstrate one dataset application example in this section. Although it would be better if a novel analysis technique could be proposed, this is outside the scope of the present paper. The provided application example is the automatically generated correlation analysis results of the dataset using MISP.

### 5.1. Noise Removal

As described in section 4, the dataset includes several types of noise, which makes a further data analysis difficult and causes erroneous results. The dataset contains noises owing to the malfunctions of the data extraction methods and the inclusion of less meaningful data. An effective noise removal technique should be able to consider the contextual necessities of the data among the entire dataset or sets of events. For example, the data contained in several sets of related events where there is little similarity of each event set is considered noise with a high probability because it increases the dissimilarity of the event sets correlated with the data.

### 5.2. Correlation Analysis

A proper usability of the dataset comes from finding the underlying relations among the data. Without any correlations, the dataset itself is nothing but a significant amount of scattered data that can only be used to search for the existence of certain items.

Because an event in the dataset is composed of several threat data, the correlations between events are determined by analyzing the relations among the threat data consisting of such events. A string-matching based method provided by many commercial cyber intelligence services would be one way to find the relations of events. However, this simple method has several limitations. If two events contain attacker names such as "Bart Simpson" and "B. Simpson", a simple string-matching based method will not find the relations between the events. Similarly, if the events include the URLs, "bartsimpson.com" and "bsimpson.net", the relations will also not be discovered. A string-similarity analysis and heuristics can be adopted to overcome such a limitation. Moreover, probabilistic approaches can improve an event-wise analysis when considering the relations among sets of data of the events.

## 5.3. Temporal Analysis

Understanding the history of cyber campaigns by adversaries is crucial, not only to defend against current incidents and presume the underlying intent, but also to draw the direction of adversarial activities from the big picture. Furthermore, the tactics, techniques, and procedures identified from the campaigns through a temporal analysis can be used to characterize the behavior of the adversarial groups. Therefore, the characteristics can be applied as a feature for a correlation analysis of the sequences of events.

## 5.4. Example Dataset Application

The proposed dataset can be used for a correlation analysis of cyber incidences. The cyber threat actor group retrieving the correlation in the example is the Lazarus group, which has been suspected to have conducted many major cyber campaigns, including the following:

- The Sony Pictures Entertainment attack (2014)
- A bank heist including the Bangladesh Bank (2016)
- The worldwide WannaCry ransomware distribution (2017)

We conducted a correlation analysis of a dataset collected by CTIMiner with help from the MISP correlation graph shown in Figure 10.

The starting point of the correlation analysis is a security report on "Lazarus' False Flag Malware [31]", marked as ⓐ. As mentioned in the report, the Lazarus group was involved in a polish banks heist, the corresponding report of which is [32] marked as ⓑ. The data-wise correlation of incidents can be found in Figure 10. The data in ①, which are extracted from the reports and from malware analysis results, correlate ⓐ and ⓑ, and the others in ② link

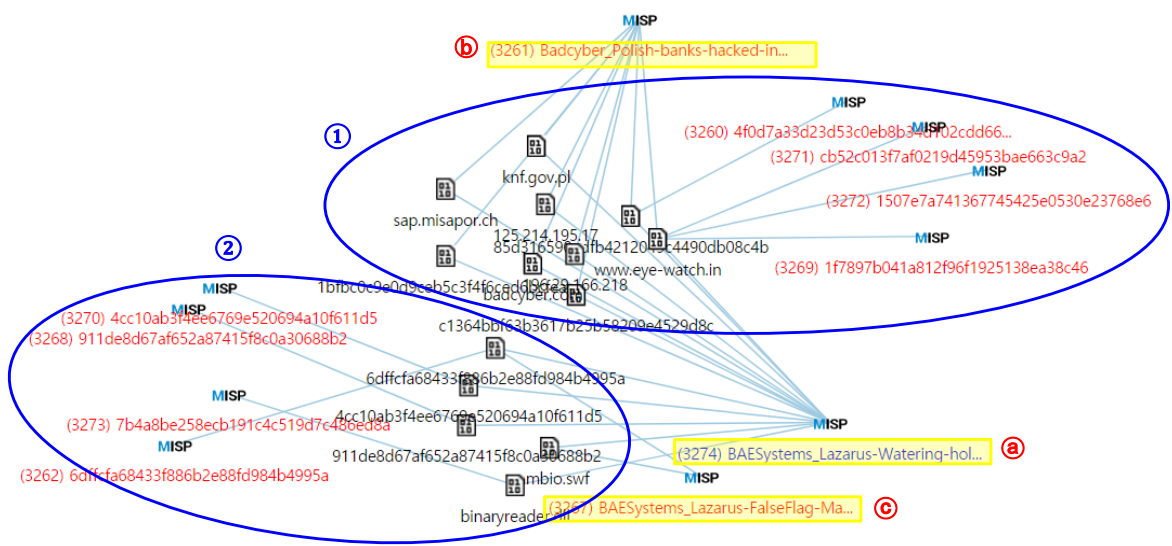

Figure 10: Sample application of the dataset to a correlation analysis

ⓐ to ©, which is another report from BAE systems regarding the Lazarus group. Therefore, through ⓐ, ⓑ, and © may have a correlation.

Although this paper does not intend to propose CTI analysis techniques, by applying a previously proposed dataset application, we can deduce practical lessons on how this dataset can be used for CTI generation in this example. A CTI analysis algorithm is basically able to find the connectivity of the data extracted from the same APT report. In advance, the algorithm can correlate the reports that analyze the same attributes and campaigns. A CTI analysis algorithm should eventually aim to generate actionable intelligence allowing patterns of attack to be determined as a means to predicting the intent of the attackers and to prepare against similar attacks.

Kim et al. proposed an event-centric correlation analysis approach to assist in generating such CTI. They suggested a novel concept and a construction algorithm that expresses the similarities among threat events and temporal characteristics in a graphical structure [12]. To use our CTI dataset for an advanced analysis, successive studies should be conducted.

## 6. Source code and Dataset Access

The source codes of the CTIMiner system and the generated dataset described in this paper are available to the public at our GitHub repository[4]. Using the source codes, security reports, and MISP, a dataset composed of the data types of interest can be generated.

## 7. Conclusion

Owing to the prevalence of cyber threats and a rapid increase in the amount of data collected, researchers are developing techniques for the different intelligence processes to be actively conducted. However, compared to other intelligence processing steps, studies have been undertaken limitedly for the analysis and production step that requires the real CTI dataset for the analysis. We pointed out that dataset unavailability is the main reason suppressing vitalization of the research despite many interests. To address the problem, we proposed CTIMiner system that generates the dataset consisted of the data contained in security reports and supplemented with malware analysis data related to the reports. After categorizing the types of data collected from the system, we provided the statistical feature of the dataset. To show the usability and applicability of the dataset, we proposed several research topics possible to be conducted using the dataset and demonstrated the correlation analysis result for an event in the dataset.

Our future research direction is to develop and enhance the proposed analysis technique using the dataset on top of the CTI correlation analysis framework [12]. By releasing this dataset to the public, we believe it can promote the threat analysis research to generate CTI.

## 8. Acknowledgement

This work was supported under the framework of international cooperation program managed by National Research Foundation of Korea (No.2017K1A3A1A17092614).

---

[4] https://github.com/dgkim0803/CTIMiner